\documentclass[aps,prx,reprint,groupedaddress,superscriptaddress,nofootinbib,longbibliography]{revtex4-2}

\usepackage{graphicx}
\usepackage{dcolumn}
\usepackage{bm}
\usepackage{hyperref}

\usepackage[export]{adjustbox}
\usepackage{graphicx}
\usepackage{dcolumn}
    \usepackage{bm}
\usepackage[utf8]{inputenc}
\usepackage{amsmath}
\usepackage{amsfonts}
\usepackage{amssymb}
\usepackage{graphicx}
\usepackage{color}
\usepackage{ bbold }
\usepackage{blindtext}

\usepackage{color}
\usepackage{ textcomp }
\usepackage{comment}

\usepackage{amssymb,amsfonts,amsmath}
\usepackage{graphicx,amsmath,bbm} 
\usepackage{subfigure}

\usepackage{amsfonts}
\usepackage[version=4]{mhchem}
\RequirePackage{amsmath,amssymb,latexsym}

\usepackage{tikz}

\usepackage{bbold}

\usepackage{siunitx}
\DeclareSIUnit\molar{\mole\per\cubic\deci\metre}
\DeclareSIUnit\Molar{\textsc{m}}
\usepackage{fancyhdr}
\fancypagestyle{style1}{
\fancyhf{}

\fancyhead[R]{S\thepage}
}

\begin{document}
\title{The prebiotic emergence of biological evolution}

\author{Charles D Kocher}
\affiliation{Laufer Center for Physical and Quantitative Biology, Stony Brook University, Stony Brook, NY 11794}
\affiliation{Department of Physics and Astronomy, Stony Brook University, Stony Brook, NY 11794}

\author{Ken A Dill}
\email{To whom correspondence may be addressed. Email: dill@laufercenter.org.}
\affiliation{Laufer Center for Physical and Quantitative Biology, Stony Brook University, Stony Brook, NY 11794}
\affiliation{Department of Physics and Astronomy, Stony Brook University, Stony Brook, NY 11794}
\affiliation{Department of Chemistry, Stony Brook University, Stony Brook, NY 11794}

\begin{abstract}
    \textbf{Abstract:} The origin of life must have been preceded by Darwin-like evolutionary dynamics that could propagate it. How did that adaptive dynamics arise?  And from what prebiotic molecules? Using evolutionary invasion analysis, we develop a universal framework for describing any origin story for evolutionary dynamics. We find that \textit{cooperative} autocatalysts, i.e. autocatalysts whose per-unit reproductive rate grows as their population increases, have the special property of being able to cross a barrier that separates their initial degradation-dominated state from a growth-dominated state with evolutionary dynamics. For some model parameters, this leap to persistent propagation is likely, not rare. We apply this analysis to the \textit{Foldcat Mechanism,} wherein peptides fold and help catalyze the elongation of each other.  Foldcats are found to have cooperative autocatalysis and be capable of emergent evolutionary dynamics.
\end{abstract}

\maketitle


\section*{Introduction}

It is not known how life arose from prebiotic matter 3.5 billion years ago.  It has not been replicated in a lab.  In the absence of experiments, there is a role for theory and modeling to help generate hypotheses.  On the one hand, there have been speculations about  ``chicken-or-egg'' questions: ``What bio-like molecules might have come first?''  Maybe life started as an \textit{RNA World} \cite{gilbertOriginLifeRNA1986,joyceProtocellsRNASelfReplication2018,atkinsRNAWorldsLife2011,readerRibozymeComposedOnly2002,horningAmplificationRNARNA2016,wochnerRibozymeCatalyzedTranscriptionActive2011,johnstonRNACatalyzedRNAPolymerization2001}; a \textit{Lipid World} \cite{segreLipidWorld2001,damerHotSpringHypothesis2020,deamerOriginsLifeCentral1994,lancet2018systems,armstrong2018replication,segrePrebioticEvolutionAmphiphilic2000,rufoShortPeptidesSelfassemble2014,greenwaldAmyloidAggregatesArise2016,takahashiConstructionChemicallyConformationally2004}; an \textit{Amyloid World} \cite{mauryAmyloidOriginLife2018,mauryOriginLifePrimordial2015,maurySelfPropagatingVSheetPolypeptide2009,routPrebioticTemplatedirectedPeptide2018,carny2005model,greenwald2010biology}; or \textit{Metabolism Came First}, where some biochemical reactions didn't require enzyme catalysts \cite{muchowskaNonenzymaticMetabolicReactions2020,wachtershauserEnzymesTemplatesTheory1988,shapiroSmallMoleculeInteractions2006,jordanSpontaneousAssemblyRedoxactive2021,codyTransitionMetalSulfides2004}.  Alternatively, the first step toward life could have involved two or more bio-like molecules~\cite{friedPeptidesNucleotideWorld2022,carterProposedModelInteraction1974,giacobelliVitroEvolutionReveals2021,willsInsuperableProblemsGenetic2018,alvaVocabularyAncientPeptides2015,dale2006protein,adamalaNonenzymaticTemplateDirectedRNA2013,pressmanRNAWorldModel2015}.  \\

We reason instead about what \textit{driving forces and dynamics} would have led to sustained bio-like propagation \cite{kocherDarwinianEvolutionDynamical2023,kocherNanoscaleCatalystChemotaxis2021,kocherOriginsLifeFirst2023,pross2003driving,dillDrivingForcesOrigins2021,prossWhatLifeHow2016}.   Why was there any tendency at all to create biology?  What process might have led polymers (such as lipids, RNA, DNA or proteins) to have specific sequences or assemblies that perform biological functions?  While physical and chemical processes tend toward equilibria and degradation according to the Second Law of Thermodynamics, biology is driven by input resources to survive, evolve, and innovate.  How did a dead regime dominated by degradation, dilution, and transience become a living regime dominated by propagation, evolution, and persistence? \\

Arguably, a Darwin-like evolutionary process must have preceded the origin of life.  As a metaphor, computers can't operate until they have an operating system.  A widely accepted definition of living system---due to NASA \cite{joyceForewardOriginsLife1994}---is that ``life is a self-sustaining chemical system \textit{capable of Darwinian Evolution}.''  The italics are ours, emphasizing the implication that since life cannot be defined in the absence of its adaptation dynamics, then some form of that dynamics must have been operating at or before the origin of life.  Life can't originate until it can propagate. This prebiotic evolution-like process could then act as the driving force that steered prebiotic chemistry toward biology \cite{prossGeneralTheoryEvolution2011,kocherOriginsLifeFirst2023,dillDrivingForcesOrigins2021}. The question of the origin of life then becomes a search for the origin of some \textit{dynamical evolutionary mechanism or process}. \\

To identify a dynamical origins mechanism requires an analysis at two levels: a macro and micro consideration.  At the macro level, we seek the broadest possible statement about what types of fluctuations occurring within an unstable degradation-dominated world could drive a transition to a stable growth-dominated world, independent of any particular microscopic model instantiation of it. At the micro level, we then seek a molecular mechanism that can satisfy this macro criterion for transition to dynamical persistence, and which also has minimal free parameters, is physical, and is prebiotically plausible. \\

We begin with the macro analysis.  We apply a universal framework called ``first invasion'' analysis that can be used to probe any proposed origin story for evolutionary dynamics, no matter what its underlying thesis in molecular physics about the origin of life.  In short, we are looking for dynamical principles of ``bootstrapping'', i.e., of how prebiotic physical and chemical processes dominated by degradation, dilution and decay could transform to stably persistent positive-feedback autocatalysis. \\

\section*{First invasion analysis shows three scenarios}

The beginnings of some form of evolutionary dynamics must have been when an autocatalyst\footnote{An agent $A$, such as a molecule, is an \textit{autocatalyst} if it accelerates the production of more of itself from some resource materials $B$, $A + B \rightarrow 2A$. For autocatalytic sets, a collection of agents $\{A\}$ can be used with only small changes to the mathematics. A \textit{cooperative autocatalyst} is, for example, one with reproductive reaction $2A + B \rightarrow 3A$, so that the mass-action reaction rate depends on higher powers of $A$ than linear.} (or autocatalytic set \cite{kauffmanAutocatalyticSetsProteins1986,hordijkHistoryAutocatalyticSets2019,kauffmanCellularHomeostasisEpigenesis1971,hordijkConditionsEvolvabilityAutocatalytic2014}) was able to establish a persistent population of itself. Evolutionary selection would then act on the variation among the characteristics of the autocatalysts and ``remember'' the best traits by enhancing them in the population via competition for resources \cite{kocherDarwinianEvolutionDynamical2023,kocherOriginsLifeFirst2023,kirschner2005plausibility}. Without a persistent population, there is no way for evolutionary dynamics to lock-in its good discoveries and commence its hallmark fitness-ratcheting process. \\

Suppose an autocatalyst is discovered by a prebiotic chemical process; what would be its fate? Would it grow into a persistent population and establish evolutionary dynamics, or would it decay away before it could do so? Evolutionary invasion analysis \cite{barabasChessonCoexistenceTheory2018,turelli1978does,chesson1989invasibility,gard1984persistence,schreiber2011persistence} is a mathematical method for determining whether an individual (the invader), when inserted into a pre-established community, will multiply or die out. The origin of evolution was the ``first successful invasion,'' the first time where a small population of autocatalysts tried to grow into an environment and succeeded. Therefore, we use an evolutionary invasion analysis to model it. \\

\begin{figure}[t]
    \centering
    \includegraphics[width =\linewidth]{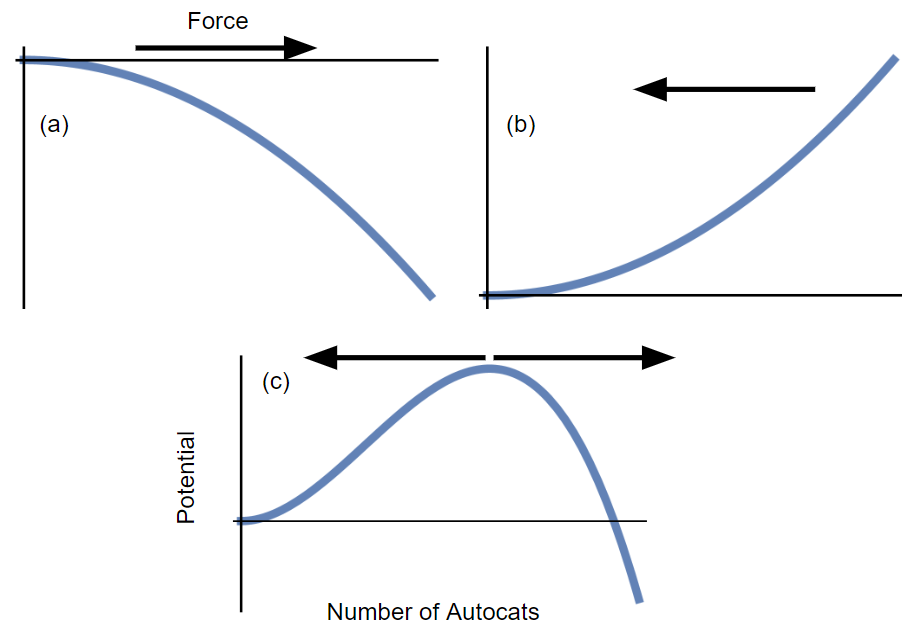}
    \caption{\textbf{Three possible potential landscapes for invasion.} (a) Introduced into a \textit{favorable} environment, the invader population grows (until it is limited by the amount of resources; see Figure~\ref{fig:finite-populations}). (b) Introduced into an \textit{unfavorable} environment, the population is pushed down to zero. (c) Degradation at low population; growth at high population.  The force changes sign at the potential minimum.}
    \label{fig:three-options}
\end{figure}

In invasion analysis, the initial population of the invading species is taken to be small enough that it doesn't perturb the existing community.  In a thermodynamics metaphor, this is like a system connected to an infinite thermal bath that it cannot change. In this unperturbing limit, the environment is fixed, so there is only a single dynamical population $A(t)$.  In the most general terms, the population will be governed by one differential equation $dA/dt = g(A)$, where the function $g$ depends on the known environmental dynamics. We assume that the timescale of environmental changes is longer than the molecular timescales; while not essential, this simplification allows us to treat $g(A)$ as a constant function (without it, we have $g(A,t)$, where the time dependence is presumed to be known). The invasion analysis limit, which we focus on here, is when $A \rightarrow 0$. We discuss various models of the full resource dependence that extend our analysis to larger values of $A$, as well as the applicability of our invasion analysis approximation, in Appendix~\ref{app:invasion-analysis}, which also outlines the types of population dynamics models that are the starting point for our analysis. \\

In general, the population $A(t)$ obeys a minimization principle, which we will use to analyze its dynamics.  If we define

\begin{equation}
    V(A) = - \int^A_0  dx \; g(x) \; , \label{eq:fitness-landscape}
\end{equation} 
then ${d V(A(t))}/{dt} = (dV/dA) (dA/dt) =  -(g(A(t)))^2 \leq 0$ and the population $A(t)$ will tend toward a value that minimizes $V(A)$ while on the path defined by $dA/dt = g(A)$. This potential landscape is a Lyapunov function for the dynamics of the autocatalyst \cite{yangPotentialsContinuousMarkov2021}. Stable steady-states are minima of $V$, because at its minima $dV/dA = -g(A) = 0$, and $d^2 V/dA^2 = -dg/dA > 0$ guarantees a restoring force ($-dV/dA$) pointing back to the steady-state value. For a population $A(t)$ to persist and undergo evolution, there only needs to be one such non-zero minimum, where it will sit indefinitely. \\

In the limit of $A \rightarrow 0$, there are only three relevant categories of potential functions, which are visualized in Fig~\ref{fig:three-options}. Any function $V(A)$ as defined in Eq~\eqref{eq:fitness-landscape}, when viewed in the small $A$ limit, will fall into one of these three categories:  \textit{(a) Favorable}.  The invader is introduced into a favorable environment and grows until it is limited by resources. The nonequilibrium-driven supply of resources sustains a force $-dV/dA$ (black arrow) that pushes the population of $A$ higher. \textit{(b) Unfavorable.}  The environment is unfavorable and the population of $A$ dies out. \textit{(c) Metastable with a tipping point.}  For the small initial population, the invader population does not grow, but for higher populations it does. There is a tipping point that is the transition from a regime of decay to a regime of persistent growth.  \\

These three behaviors are expressed by a general Taylor-expanded version of the ODE for $A(t)$:
\begin{equation}
    \frac{dA}{dt} = g(A) \approx (g_1 - D) A + g_2 A^2 \; . \label{eq:templated-polymerization-type}
\end{equation}
where $g_1$ is a growth rate, $D$ is a decay or degradation rate, and $g_2$ is the rate coefficient for a lowest-order nonlinear cooperativity effect. These terms are all that are needed to capture the three fates of the autocatalyst population shown in Fig~\ref{fig:three-options}. The corresponding potential function of this simple model is
\begin{equation}
V(A) = \frac{(D - g_1) A^2}{2} - \frac{g_2 A^3}{3} \; .  \label{eq:invasion-analysis}
\end{equation} 
Case (a) is when $g_2 \geq 0$ and $g_1 > D$, case (b) has $g_2 \leq 0$ and $g_1 < D$, and case (c) has $g_2 > 0$ and $g_1 < D$. Case (c) requires a \textit{cooperative autocatalyst}. Here, we are defining cooperativity in the same way it is defined in binding polynomials. For example, hemoglobin binds to a second oxygen ligand more tightly when a first oxygen is already bound to it \cite{hill1910possible,adair1925hemoglobin,abeliovich2005empirical}. In our situation, cooperativity means the autocatalyst gets better at making itself as its population goes up: the birth rate is $(g_1 + g_2 A)A$, where both constants are positive. Our term cooperativity is a shorthand for positive cooperativity, i.e. where $g_2 > 0$, not negative cooperativity. We note that cooperativity is not a guaranteed property of any autocatalyst that prebiotic chemistry could have discovered: Appendix~\ref{app:invasion-analysis} discusses further when autocatalysts are considered cooperative for the purposes of the invasion analysis approximation. Case (c) cannot be realized by non-cooperative invaders. Case (b) is non-cooperative ($g_2 = 0$) or negatively cooperative ($g_2 < 0$), and the dynamics of case (a) do not change if the positive cooperativity is removed because the first term of the Taylor expansion is sufficient. We should note here that $g_1$, $g_2$, and $D$ are constants because of our assumption that environmental changes are slow; when environmental changes matter, each of these just becomes a (known) function of time, so that the potential landscape that the autocatalyst sees can change. It is only the current potential landscape that determines the autocatalyst's behavior, since the ODE dynamics are first order in time. We emphasize again that no matter the underlying mechanism or origin of life model of the invading autocatalyst, these three cases can be applied: they are the only possible behaviors. \\

The deterministic behavior of case (a) of Fig~\ref{fig:three-options} is to give a persistent population of autocatalysts undergoing evolutionary dynamics, while cases (b) and (c) predict that the autocatalysts will die out and there will be no evolution. However, the real dynamics are not deterministic. Cases (a) and (b) do not change when noise is added, but case (c) does. The population $A(t)$ will move around stochastically about its deterministic path. Furthermore, the environment itself is fluctuating. Mathematically, this manifests as fluctuations in the quantities $g_1$, $g_2$, and $D$ of Eq~\eqref{eq:templated-polymerization-type}. Graphically, this means that the location of the peak of the potential barrier of Fig~\ref{fig:three-options}(c) itself can shift. Through these combined motions, case (c) autocatalysts can hop the potential barrier between decay (left side of the barrier) and growth (right side of the barrier). Once it is on the right side, the population will have a sustained driving force toward even higher population levels far away from the barrier, establishing a persistent population that is able to evolve. So, cases (a) and (c) can describe an origin of evolution, while case (b) cannot. 

\subsection*{Using first invasion analysis to probe origin stories.}

The first invasion analysis described above is completely general, capable of assessing any particular model of origins of life, as we will now argue. It encompasses previous approaches applied to specific models of prebiotic RNA templated polymerization \cite{wuOriginLifeSpatially2012,shay2015origin} and of a prebiotic disorder-to-order transition \cite{dysonModelOriginLife1982,dysonOriginsLife1999}.  Our main specific application of the first invasion analysis framework will be to the Foldcat Mechanism of peptides in the next section, but in Appendix~\ref{app:invasion-analysis} we also illustrate its applicability to a simplified model of templated polymerization. \\

For evolution to emerge from prebiotic chemistry, an autocatalyst must eventually be discovered that is case (a) or case (c). There is no way to begin evolutionary dynamics without discovering the autocatalyst that can evolve (the driving force needs a medium to act on), and the only ways to introduce it are via cases (a) and (c). Origin stories that fit case (a) are of the ``right place at the right time'' nature. Prebiotic chemistry would have discovered an autocatalyst in an ideal environment that favored its growth over decay.  The origin story must then explain how that perfect match between environment and autocatalyst was produced using only prebiotic chemistry. \\

We are interested in mechanisms that are case (c) because these cooperative autocatalysts are able to cross the potential barrier to growth even in unfavorable environments. Moreover, as demonstrated in the specific model of \cite{wuOriginLifeSpatially2012,shay2015origin}, this event only needs to happen in one spatially localized place, from which the autocatalysts can diffuse elsewhere into other environments, causing potential barrier crossings wherever they go. Autocatalysts that are non-cooperative, that is they are case (a) in some places and case (b) in others, cannot do this. Adding more autocatalysts to a case (b) environment cannot flip it into a self-sustaining population; it will always require diffusion from the favorable environment. Diffusion of autocatalysts into a case (c) environment causing a barrier hopping, however, can ignite a self-sustaining population of the autocatalyst that no longer relies on the diffusion of autocatalyst inward. Thus, case (c) has an ``any place at any time'' nature.  If a cooperative autocatalyst is repeatedly re-introduced into the same case (c) environment, it will \emph{inevitably} hop the potential barrier if given enough time and attempts (depending on the barrier hopping probability and the rate of re-introduction, this amount of time could be unphysically long, however). Think of it as a biased coin-flip, where the probability of heads is the non-zero probability of hopping the potential barrier. Given enough flips, there will eventually be a heads. Cooperative autocatalysts have a chance, or even a likelihood, of survival even in very poor environments. Cooperativity allows for a much broader range of conditions for the emergence of evolution. 

\section*{The Foldcat Mechanism and its emergent evolutionary dynamics}

\begin{figure}[t]
    \centering
    \includegraphics[width = \linewidth]{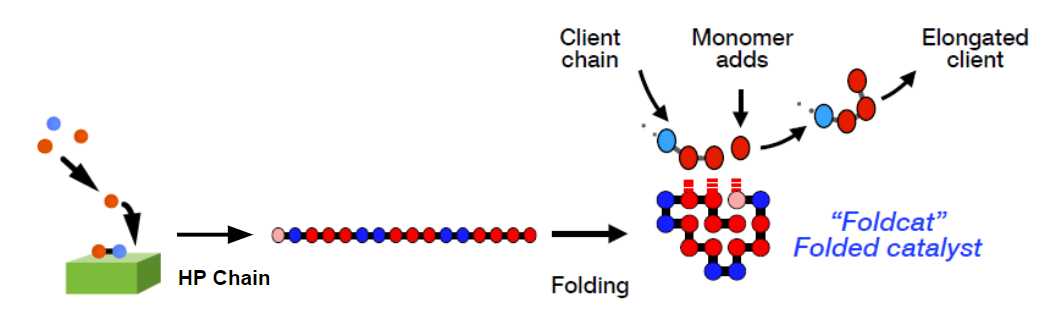}
    \caption{\textbf{The Foldcat Mechanism.} Short peptides of hydrophobic (H) and polar (P) monomers are synthesized on the ``Founding Rock'' catalyst (green).  A few long chains fold to stable structures with catalytic-competent surfaces.  These can elongate other chains, giving positive feedback of chain-length growth.}
    \label{fig:stationary-to-foldcat}
\end{figure}

We now give a model at the micro level. The \textit{Foldcat Mechanism}, described previously \cite{gusevaFoldamerHypothesisGrowth2017,kocherDarwinianEvolutionDynamical2023,kocherOriginsLifeFirst2023,dillDrivingForcesOrigins2021}, postulates that the synthesis of random short-chain peptides leads to small populations of longer protein molecules that can both \textit{fold} and \textit{cat}alyze chemical reactions \cite{friedPeptidesNucleotideWorld2022,matsuoProliferatingCoacervateDroplets2021,ikeharaGADVProteinWorld2014,frenkel-pinterPrebioticPeptidesMolecular2020,leeSelfreplicatingPeptide1996,kauffmanAutocatalyticSetsProteins1986}; see Fig~\ref{fig:stationary-to-foldcat}.  Short random HP (hydrophobic/polar, referring to the two types of monomers) peptides are synthesized, catalyzed at first through some macroscale object, like clays or minerals which, as shorthand, we call \textit{the Founding Rock}.  A small fraction of those chains are longer, collapsing into compact conformations with a hydrophobic core.  Some of these stable folders have reactive surfaces.  Indicated here as having hydrophobic sticky ``landing pads,'' these chains act as catalysts that grab other peptides and hydrophobic monomers in juxtaposition, accelerating elongation of the client chain (by up to several orders of magnitude according to some estimates \cite{mengerInteractionVsPreorganization2019}). We call chains that fold and catalyze elongation \textit{foldcats}.  Here, we analyze the Foldcat mechanism by invasion analysis to ask if that mechanism admits of parameters that could enable the disorder-to-order bootstrapping transition needed for the origins of evolution. \\


A key question in the origins of life is what \textit{fitness} might have been before there were cells, in a world that contained only molecules.  Through what mechanisms or actions could molecules become self-serving?  In the Foldcat Mechanism, the \textit{fitness ratcheting} that selects winners from losers is simply molecular \emph{persistence} in the environment. Polymer chain sequences that fold more stably will survive longer.  And, chains that are autocatalytic (helping to elongate others) drive to further increase the populations of molecules that are long and autocatalytic.  \\

There has not been a direct experimental test of the Foldcat Mechanism, but there is experimental evidence for the ideas of folding persistence as the first evolutionary driving force \cite{shibueComprehensiveReductionAmino2018,makarovEarlySelectionAmino2023,makarovEnzymeCatalysisPrior2021,riddleFunctionalRapidlyFolding1997,kimuraReconstructionCharacterizationThermally2020,longoSingleAromaticCore2015,yagiSevenAminoAcid2021} and hydrophobic amino acids driving peptide ligation \cite{fodenPrebioticSynthesisCysteine2020,routPrebioticTemplatedirectedPeptide2018,takahashiConstructionChemicallyConformationally2004}. Amino acids and short peptides have been generally regarded as existing on the early earth \cite{millerProductionAminoAcids1953,millerOrganicCompoundSynthesis1959,friedPeptidesNucleotideWorld2022,parkerPrimordialSynthesisAmines2011,johnsonMillerVolcanicSpark2008,croninAminoAcidsMeteorites1983,glavinExtraterrestrialAminoAcids2021,kebukawaGammaRayInducedAminoAcid2022,holdenAqueousMicrodropletsEnable2022,griffithSituObservationPeptide2012,dealWaterAirInterfaces2021,furukawaAbioticFormationValine2012,lambertAdsorptionPolymerizationAmino2008,takahagiPeptideSynthesisAlkaline2019}.  And, reasons have been given for why the most likely first steps entailed proteins, or proteins plus RNA, and not RNA alone \cite{friedPeptidesNucleotideWorld2022,kitadaiOriginsBuildingBlocks2018,willsInsuperableProblemsGenetic2018,dillDrivingForcesOrigins2021,frenkel-pinterPrebioticPeptidesMolecular2020}.

\subsection*{Two cooperativities: Folding slows degradation.  Catalysis accelerates elongation.} This mechanism entails two contributions to autocatalytic cooperativity; namely, that folded chains degrade slower than unfolded ones because they have protected cores, and that some foldcats serve as catalysts to accelerate the production of longer chains.  This mechanism bootstraps to produce longer, more folded, more catalytic molecules. Because of their cooperative feedback, the more foldcats that arise, the higher the rate of producing long, stable, catalytically active chains. \\  

\begin{figure*}[t]
    \centering
    \includegraphics[width = 0.9\linewidth]{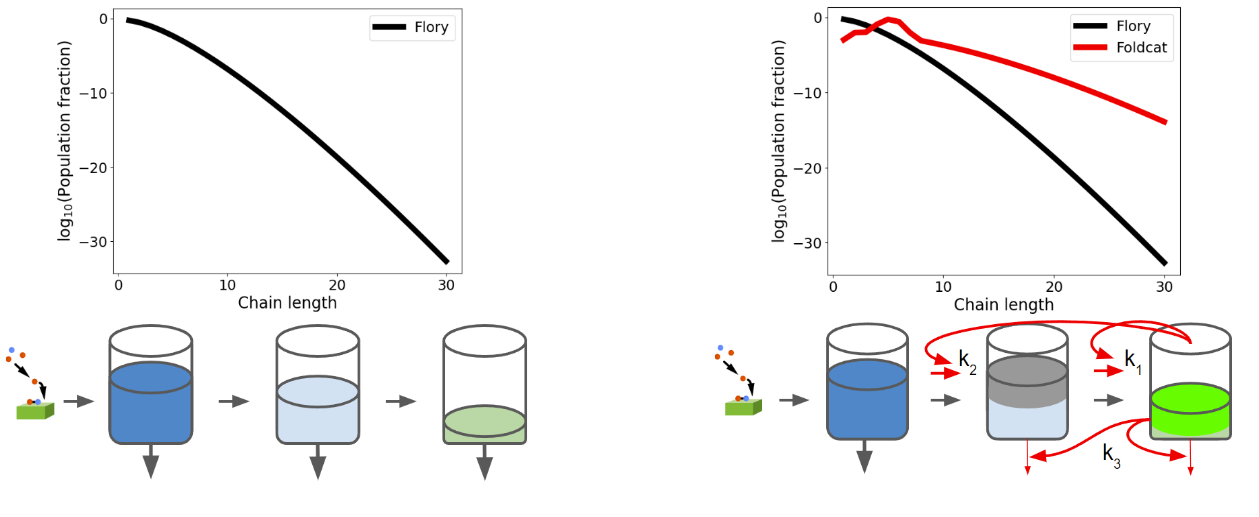}
    \caption{\textbf{Chain length distributions created by foldcats' cooperativity.}  (Left, black) Typical polymerizations produce mostly short chains (shown as bin filling), following a Flory distribution \cite{flory1953principles,gusevaFoldamerHypothesisGrowth2017}. (Right, red) Foldcats produce more long chains for 2 reasons:  (i) Foldcats catalyze elongation of other chains ($k_1$ and $k_2)$, and (ii) Folded chains degrade more slowly because their folded cores are protected from solvent ($k_3$). For analytical forms, see Appendix~\ref{app:full-foldcat-model}. }
    \label{fig:guseva-buckets}
\end{figure*}

Fig~\ref{fig:guseva-buckets} shows the two cooperativity factors of the Foldcat Mechanism.  The left figure is the baseline model powered by the Founding Rock and a supply of monomers:  it makes many short chains, fewer medium-length chains, and even fewer long chains. The concentrations of monomers flowing into each chain bin (short, medium, long, folder, foldcat, etc) can be visualized by the filling of buckets. Each bucket drains into the next bucket on the right, and its chains degrade out of the bottom of the bucket at a fixed rate. The resulting population distribution is plotted above the buckets. The right figure shows the speed-ups that the Foldcat Mechanism provides beyond the Founding Rock: (1) Foldcats (bucket three) elongate chains that are their direct precursors (bucket 2) into foldcats with a rate $k_1$, (2) Foldcats elongate \emph{the precursors of their precursors} (bucket 1) into the foldcat precursors (bucket 2) with rate $k_2$, and (3) The buckets to the right degrade slower because these chains are longer and more folded, with the rate of slowing related to $k_3$. The first of these three activities is regular autocatalysis; the latter two are cooperative. The result of the foldcat enhancements is the red population distribution, which gives orders of magnitude more long chains when compared to just the Founding Rock's distribution. Analytical forms of both of these population distributions, as well as the example parameters used, are given in Appendix~\ref{app:full-foldcat-model}. 

\begin{figure}[t]
    \centering
    \includegraphics[width = \linewidth]{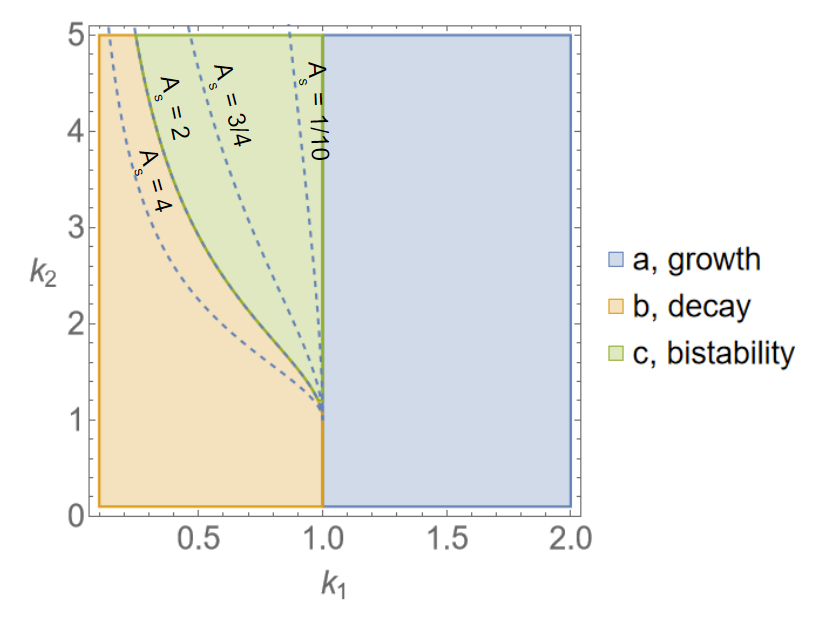}
    \caption{\textbf{Phase diagram of cases (a) growth, (b) decay, and (c) bistability for the Foldcat Mechanism model Eqs~\eqref{eq:foldcat-type} and \eqref{eq:foldcat-type-fitness}} when varying the two dimensionless parameters $k_1$ and $k_2$. The cases refer to those shown in Fig~\ref{fig:three-options}. The bistability-decay boundary is shown for various values of the free parameter $A_s$, which characterizes the concentration of foldcats at which the catalytic cooperativity starts to saturate (see Appendix~\ref{app:foldcat-model} for details). }
    \label{fig:foldcat-type-phase-diagram}
\end{figure}

\subsection*{Dynamical model of the Foldcat Mechanism.}

To keep things simple and focus on how a population of adaptive foldcats could establish itself via the ``first invasion'' framework, we do not include chain sequence or chain length information in our first model presented here (although we do include some chain length and sequence information in our models of Appendices~\ref{app:full-foldcat-model} and \ref{app:binned-model}, to be discussed more later). Our goal is to demonstrate how a foldcat-like autocatalyst with the types of cooperative feedback illustrated in Fig~\ref{fig:guseva-buckets} gives a region of case (c) metastability behavior. The basic reactions in our model are that monomer $M$ is supplied at a rate $\alpha_M$ and decays at a rate $d_M M$, while non-foldcat chains $r$ are created at a rate $\alpha_r$ and decay at a rate $d_r r$. Then, we have the elongation reactions, which are catalyzed both by the Founding Rock and by our foldcats $A$: (1) $r + M \rightarrow A$ (non-foldcat is elongated into a foldcat), (2) $r + M \rightarrow r$ (non-foldcat is elongated and still is not a foldcat),  (3) $A + M \rightarrow A$ (foldcat is elongated and is still a foldcat), and (4) $A + M \rightarrow r$ (foldcat is elongated and is no longer a foldcat). Elongation reaction $i$ has mass-action rate constant $K_i(A)$, which has a Founding Rock and foldcat contribution. The full set of differential equations describing the Foldcat Mechanism is

\begin{align}
    \frac{dr}{dt} &= \alpha_r - d_r r + K_4(A) A M - K_1(A) r M \; , \nonumber \\
    \frac{dM}{dt} &= \alpha_M - [d_M + r (K_1(A) +  K_2(A)) + \nonumber \\ 
    &+ A (K_3(A) +  K_4(A))] M \; , \label{eq:foldcat-type-full-system-ODE-MT} \\
    \frac{dA}{dt} &= K_1(A) r M - K_4(A) A M - D A \; . \nonumber 
\end{align}
After a few simplification steps (see Appendix~\ref{app:foldcat-model} for details), Eq~\eqref{eq:foldcat-type-full-system-ODE-MT} becomes a single equation in $A(t)$,

\begin{equation}
    \frac{dA}{dt} = \frac{k_1 A}{1 + k_1 A} + \frac{k_1 k_2 A^2}{(1 + k_1 A)(1 + A/A_s)} - A \; . \label{eq:foldcat-type}
\end{equation}

\noindent All variables are now dimensionless. The parameters of this mechanism are the non-cooperative reproduction rate $k_1$, which is the rate at which foldcats elongate non-catalytic chains into foldcats; the cooperative reproduction rate $k_2$, which is the rate at which foldcats create their direct precursors (a foldcat minus one monomer) from monomers or shorter non-catalytic chains; and an additional free parameter $A_s$, which is the Michaelis (saturation) constant of the creation of direct precursors from monomers or shorter non-catalytic chains. As described in Appendix~\ref{app:foldcat-model}, $k_1$ is a re-defined, dimensionless version of the parameter $K_1$, and $k_2$ arises as part of the function $\alpha_r$. The parameters $k_1$ and $k_2$ act as visualized in the bucket metaphor of  Fig~\ref{fig:guseva-buckets}. In terms of the first invasion analysis parameters of Eq~\eqref{eq:templated-polymerization-type}, Taylor expanding both of the first two terms of Eq~\eqref{eq:foldcat-type} gives $D = 1$ (definition of the dimensionless time parameter, see Appendix~\ref{app:foldcat-model}), $g_1 = k_1$, and $g_2 = k_1 k_2 - k_1^2$. Surprisingly, even if $k_2 > 0$, there is a region of negative $g_2$. Even though the nature of the cooperativity may seem straightforward, the range of parameters for which the system is cooperative may be unexpected, and a full analysis like that of Appendices~\ref{app:invasion-analysis} and \ref{app:foldcat-model} is needed. \\

The corresponding potential function for the Foldcat Mechanism Eq~\eqref{eq:foldcat-type} is

\begin{align}
    V(A) = &\frac{A^2}{2} - A - A A_s k_2 - \frac{A_s^3 k_1 k_2 \ln(1 + A/A_s)}{1 - A_s k_1} - \nonumber \\
    & - \frac{(1 - A_s k_1 + A_s k_2) \ln(1 + k_1 A)}{k_1(A_s k_1 - 1)} \; .  \label{eq:foldcat-type-fitness}
\end{align}

\noindent Using the three classes of behavior from Fig~\ref{fig:three-options}, we can create a phase diagram for foldcats from Eq~\eqref{eq:foldcat-type-fitness}; see Fig~\ref{fig:foldcat-type-phase-diagram}. The interpretation of this phase diagram is as follows: first, nature discovers foldcats in some environment; then, the values of the parameters $k_1$, $k_2$, and $A_s$ are computed for that environment, putting foldcats at one fixed point on the phase diagram; finally, the region in which the point falls determines the foldcats' fate. Each case from Fig~\ref{fig:three-options} is represented: one region predicts pure growth (blue), one is pure death (yellow), and one is bistability (green). How quickly the creation of foldcat precursors saturates (magnitude of $A_s$) determines the extent of the Fig~\ref{fig:three-options}(c) metastability region of foldcat discovery. \\

The simple model of Eq~\eqref{eq:foldcat-type} demonstrates the \textit{cat} part of the Foldcat Mechanism's cooperativity. The other form is the \textit{fold} part. Since our Eq~\eqref{eq:foldcat-type} does not have sequence or chain length information, this type of cooperativity has to be put in by hand (but it arises natively in the more detailed models of Appendices~\ref{app:full-foldcat-model} and \ref{app:binned-model}). To see the \textit{fold} cooperativity, we should change the decay term in Eq~\eqref{eq:foldcat-type}:

\begin{equation}
    -A \rightarrow -\left(A - \frac{k_3 A^2}{1 + A^2/B_s^2} \right) \; , \label{eq:foldcat-decay-cooperativity}
\end{equation}
where $k_3$ relates to the magnitude of the foldcats' ability to decrease the decay constant, and $B_s$, like $A_s$ above, characterizes the saturation of the foldcats' degradation fighting effect. The parameter $k_3$ is visualized in Fig~\ref{fig:guseva-buckets}. Note that the saturating parameters $A_s$ and $B_s$ are necessary: without them, our model would not capture all of the foldcats' possible dynamics. In this sense, the model of foldcats we put forth here is the minimal one (only one additional parameter is needed for each form of cooperativity) that captures the foldcats' type of bootstrapping physics shown in Fig~\ref{fig:guseva-buckets}. In the terms of the invasion analysis parameters of Eq~\eqref{eq:templated-polymerization-type}, adding decay cooperativity only changes $g_2$ to $g_2 = k_1 k_2 - k_1^2 + k_3$. Also, $k_3$ and $B_s$ have a constraint that the term in parentheses in Eq~\eqref{eq:foldcat-decay-cooperativity} must always be positive. This requires that $B_s^2 k_3^2 < 4$. The main effect of the decay cooperativity will be to increase the size of the bistability region in Fig~\ref{fig:foldcat-type-phase-diagram}, cutting further into the yellow case (b) decay region. The full potential function for foldcats with both cooperative catalysis ($k_2$) and cooperative chain stability ($k_3$) is given in Appendix~\ref{app:foldcat-model}. 

\subsection*{Persistence drives the emergent evolution of the Foldcat Mechanism.}

\begin{figure}[t]
    \centering
    \includegraphics[width = 0.8\linewidth]{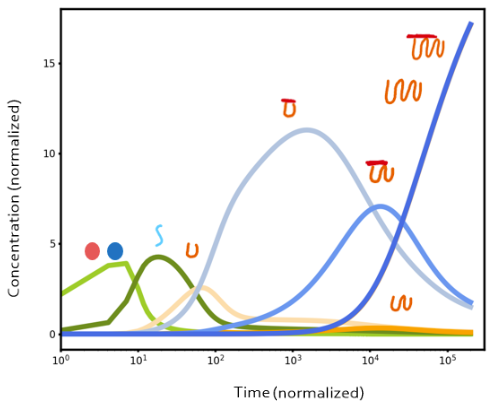}
    \caption{\textbf{Foldcat Mechanism gives evolutionary dynamics.} At first, monomers (light green) are polymerized into short random unfolded chains (dark green), then short folders (light orange) and short foldcats (pale blue), then to longer folders and foldcats (darker blue).}
    \label{fig:guseva-persistence-ratchet}
\end{figure}

To show that our conclusions about the Foldcat Mechanism from Eq~\eqref{eq:foldcat-type} do not change when information about the sequences or chain lengths are added back in, and to obtain information about the time-dependent dynamics of the Foldcat Mechanism, we simulated a modified version of our Foldcat Mechanism model which lumped chains into length and sequence ``bins.'' Fig~\ref{fig:guseva-persistence-ratchet} shows the computed time course of this modified model for the parameters given in Appendix~\ref{app:binned-model}.  The model predicts a series of epochs: first to appear are short random peptides; later are longer chains, which are enriched in folders and foldcats.  Almost all the early production are short useless peptides that degrade back to monomers.  It's a dynamical process in which small seedlings of order arise from a sea of disorder, much like modern evolutionary dynamics.  Most early molecules are random, short and unproductive.  Incrementally advantageous molecules rarely arise within this large sea of options, but when they do further advantages follow from them, and so on, until ultimately a large global advantage has been built up. Throughout the process, the ``persistence'' of chains---that is, their fold stability and elongation activity---continually increases. Persistence acts as the fitness for this evolution-like dynamics. Interestingly, we note that the searching behavior demonstrated in Fig~\ref{fig:guseva-persistence-ratchet} is also similar to a previously studied model of protein folding itself as local first, global later \cite{rollinsGeneralMechanismTwoState2014}. It's a two-step discovery process: first is random search by the Founding Rock, which is then superseded by a driven search by the foldcats.  \\

In this particular simulation, parameters were chosen so that the Foldcat Mechanism was in the case (a) region of Fig~\ref{fig:foldcat-type-phase-diagram}. This binned model does have the same cooperativities visualized in Fig~\ref{fig:guseva-buckets}, so cases (b) and (c) exist as well. In decay scenarios, the light blue ``small foldcat'' curve would stay near zero concentration, and the dynamics would stop with the first epoch generated by the Founding Rock search step. However, the Founding Rock would continually rediscover foldcats with some small rate. Since the foldcats are cooperative, if stochastics were taken into account, the small population of foldcats would fluctuate, possibly leading to the explosion in foldcats demonstrated in Fig~\ref{fig:guseva-persistence-ratchet} and the later epochs that followed. If the Founding Rock was given enough time to act, the probability of foldcats jumping the potential barrier would approach unity, meaning that the emergence of evolution is a likely property of the Foldcat Mechanism.

\section*{Conclusions} 
The question we have raised in this paper is how prebiotic non-catalytic degradation-prone reactions could have transitioned to autocatalytic persistent growth processes toward biology.  Of necessity, this requires explaining the molecular bases of cooperativities and their bootstrapping origins from simpler processes. Our invasion analysis here elucidates the macro constraints that a micro model must satisfy. But simply choosing macro parameters that predict a transition would not be an explanation of origins.  An explanation of origins requires a plausible microscopic model that has a physical basis in molecular physics,  minimal parameters, and tenable grounding in prebiotic processes.  The Foldcat hypothesis is found to satisfy these criteria.  As random peptides grow longer, they fold, protecting their cores from degradation, and they catalyze the elongation of other chains, accelerating further growth of the population of peptides.  It gives a plausible basis for the origins of biological evolution.


\section*{Acknowledgments}

We are grateful to the Laufer Center for Physical and Quantitative Biology at Stony Brook and to the John Templeton Foundation for financial support (grant ID 62564). \\




\appendix






\section{Analytical theory of the Foldcat Mechanism.}\label{app:full-foldcat-model}

\renewcommand{\theequation}{A\arabic{equation}}
\setcounter{equation}{0}

The Foldcat Mechanism has previously been explored through computational simulations \cite{gusevaFoldamerHypothesisGrowth2017,kocherOriginsLifeFirst2023}.  Here, we give an analytical approximation to it.  Assume we have a monomer $u_1$---supplied at rate $\alpha$ and decaying at rate $D u_1$--and that polymerization is able to occur, $u_1 + u_j \rightarrow u_{j+1}$ with rate $k u_1 u_j$. We will not track all of the individual rates of chains breaking apart; instead, we will just assume that there is some rate $d(n) u_n$ that chains completely fall apart (by assuming that chains just decay to nothing, we are getting a lower bound on the number of chains at each level). An ODE model of this reaction system is

\begin{align}
    \frac{du_1}{dt} &= \alpha - D u_1 - 2 k u_1^2 - \sum\limits_{n=2}^\infty k u_n u_1 \; , \label{eq:random-poly} \\
    \frac{du_n}{dt} &= k u_{n-1} u_1 - k u_n u_1 - d(n) u_n \; . \nonumber
\end{align}

Solving Equation~\eqref{eq:random-poly} is not easy in the general case, but we can use some tricks at steady-state. Interpreted the usual way, $u_1$ would be a function of all of the parameters of the problem: $\alpha$, $D$, $k$, etc. However, we can instead say that $u_1$ at steady-state is known and solve for the corresponding $\alpha$ using the rate equation for $u_1$. It is then easy to solve for each $u_n$ recursively: 

\begin{equation}
    u_n = \frac{k u_1}{k u_1 + d(n)} u_{n-1} \; . \label{eq:random-poly-un-recursive}
\end{equation}

\noindent We can subsequently rewrite $u_n$ non-recursively using a product:

\begin{equation}
    u_n = u_1 \prod\limits_{j=2}^n \left( \frac{k u_1}{k u_1 + d(j)} \right)\; . \label{eq:random-poly-un}
\end{equation}

For any set $d(j) > 0$, $u_1$ is the max of the distribution in Equation~\eqref{eq:random-poly-un}, and the population of chains falls off with increasing length. The Flory distribution \cite{flory1953principles}, 

\begin{equation}
    F_n = a (1-a)^{n-1} \; , \label{eq:flory}
\end{equation}

\noindent gives the fraction $F_n$ of chains of length $n$ and is characterized by the parameter $a$ which is the probability that one of a monomer's two connections is the end of the chain it is in. What values of the decay constants $d(j)$ would turn our distribution $u_n$ with a given $u_1$ into the Flory distribution with a given $a$ and total population $U$? Setting $U = \sum u_n$, so that we are looking for the form $u_n = U F_n$, we can divide consecutive terms $u_{n+1}/u_n = F_{n+1}/F_n$, to find

\begin{equation}
    d(n) = k u_1 \left( \frac{a}{1 - a} \right) \; , \label{eq:flory-decay-constants}
\end{equation}

\noindent a constant. If we plug this result back into the distribution $u_n$, we find that

\begin{equation}
    u_n = u_1 (1-a)^{n-1} \; . \label{eq:un-flory}
\end{equation}

\noindent We see that $U = u_1 / a$, as it should. Since the Flory distribution is a special case of our $u_n$ for a given choice of decay constants, we call Equation~\eqref{eq:random-poly-un} the generalized Flory distribution. \\

If we want to know the fraction of all monomers in chains of length $n$ in the Flory model (equivalent to the mass or weight distribution), we get

\begin{equation}
    F_{n,\text{weight}} = n a^2 (1-a)^{n-1} \; . \label{eq:flory-total-monomer}
\end{equation}

\noindent To find the total amount of monomers in chains of length $n$ in the generalized Flory model, we just multiply in an $n$:

\begin{equation}
    u_{n,\text{weight}} = n u_1 \prod\limits_{j=2}^n \left( \frac{k u_1}{k u_1 + d(j)} \right)\; . \label{eq:random-poly-un-total-monomer}
\end{equation}

\subsection*{Folding enhances the fraction of large chains}

The Flory distribution is exponentially suppressed at large chain length. What about polymers that can fold into stable configurations? For simplicity, we will consider the reaction $u_n \rightleftharpoons f_n$, with folding rate $k_f(n) u_n$ and unfolding rate $k_u(n) f_n$. We will assume that elongation only happens to unfolded chains. Making this assumption, the monomer rate equation does not change, so we find

\begin{align}
    \frac{du_n}{dt} &= k u_1 (u_{n-1} - u_n) - d(n) u_n + k_u(n) f_n - k_f(n) u_n \; , \label{eq:random-poly-plus-folding} \\
    \frac{df_n}{dt} &= k_f(n) u_n - k_u(n) f_n - d_f(n) f_n \; . \nonumber
\end{align}

\noindent To compare to the Flory distribution, we put $f_n$ to steady-state:

\begin{align}
    \frac{du_n}{dt} &= k u_1 (u_{n-1} - u_n) - \left(d(n) +  \frac{d_f(n) k_f(n)}{k_u(n) + d_f(n)} \right) u_n \; , \label{eq:folding-steady-state-assumption} \\
    f_n &= \frac{k_f(n)}{k_u(n) + d_f(n)} u_n \; . \nonumber
\end{align}

\noindent The form of the $u_n$ rate equation is now the same as in the case without folding, where only the decay constants have been modified. The total amount of polymer is 

\begin{equation}
    p_n = u_n + f_n = \left(1 + \frac{k_f(n)}{k_u(n) + d_f(n)}\right) u_n \; . \label{eq:folding-total-polymer}
\end{equation}

\noindent Using the generalized Flory distribution, we can plug in to find

\begin{equation}
    p_n = \left(1 + \frac{k_f(n)}{k_u(n) + d_f(n)} \right) u_1 \prod\limits_{j=2}^n \frac{k u_1}{k u_1 + d(j) + \frac{d_f(j) k_f(j)}{k_u(j) + d_f(j)}} \; . \label{eq:pn-general-case-folding}
\end{equation}

We should note that this $p_n$ cannot be directly compared to the generalized Flory distribution because the mapping between $u_1$ and $\alpha$ is different. The correct comparison would be to look at the two with the same $\alpha$, not the same $u_1$. However, in the realistic case where $d_f(n) \ll d(n)$, we can ignore its contribution to the $u_n$ rate equation. The result is that $u_n$ is exactly the generalized Flory distribution result, and that $p_n = \left(1 + \frac{k_f(n)}{k_u(n)}\right) u_n$. Here, the mapping between $u_1$ and $\alpha$ is the same, so we can directly compare. The enhancement over the generalized Flory distribution depends solely on the ratio of folding to unfolding, which can increase the population of polymer by orders of magnitude. 

\subsection*{Foldcats further beat the Flory distribution}

\begin{figure}[t]
    \centering
    \includegraphics[width = \linewidth]{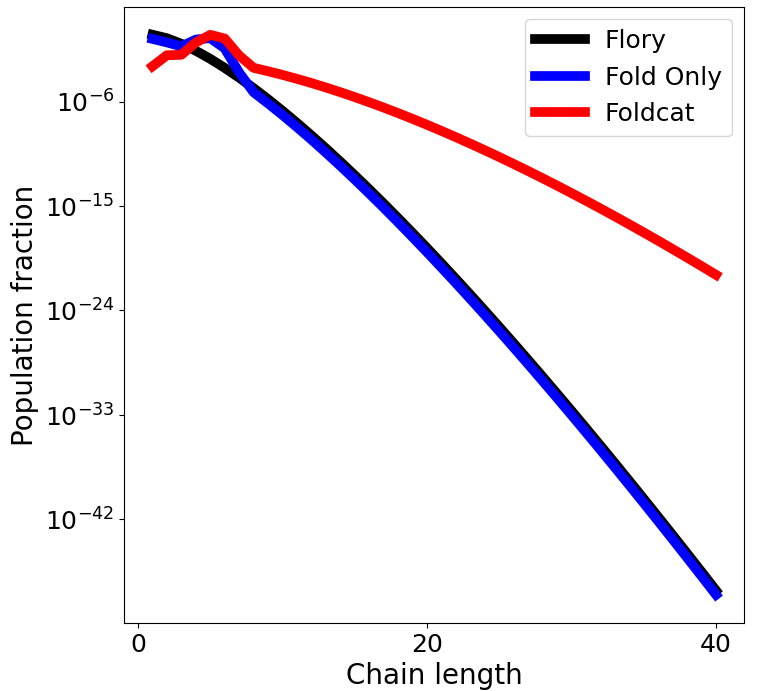}
    \caption{\textbf{Foldcats enrich long chains by orders of magnitude.} Population fraction as a function of chain length for the generalized Flory mechanism (Eq~\eqref{eq:random-poly-un-total-monomer}, black curve), the elongation of foldable chains (Eq~\eqref{eq:random-poly-plus-folding}, blue curve), and the elongation of foldcats (Eq~\eqref{eq:foldcat-full}, red curve). The total amount of chains increases as folding and catalysis are added, as well as the fraction of long chains produced. Parameters used for this plot are the same as the ones used in Fig~\ref{fig:guseva-buckets} in the main text.}
    \label{fig:random-fold-foldcat}
\end{figure}

We now introduce foldcats, as in \cite{gusevaFoldamerHypothesisGrowth2017}. As we mentioned before, we will not track polymer sequence here. Instead, we will continue to use only length information in our ODEs. Unlike in \cite{gusevaFoldamerHypothesisGrowth2017}, we are thus not limited in the number of polymers we can track. To deal with the issue of tracking the exponentially growing number of sequences as a function of chain length, Guseva et al. introduced a dilution term that kept the total number of chains manageable for their simulation. This dilution completely wiped out the enhancement of folding alone that we reported above after simplifying Eq~\eqref{eq:pn-general-case-folding}. In our model, we can keep the decay rates general. \\

\begin{center}
\begin{table}[t!]
\begin{tabular}{ ||c | c||} 
 \hline
  Concentration units & $u_1 = 1$ \\
  Time units & $1/d(2) = 1$  \\  
  $d(n)$ & $d(2) + (n - 2)$ \\
  $k$ & 1\\ 
  $d_f(n)$ & 0 \\
  $k_f(n)$ & $\exp\left[-(n-5)^2\right]$  \\
  $k_u(n)$ & $\exp(-n) + \exp(n-13) $ \\
  $k_c$ & 1.8  \\  
  $C$ & 4.145 \\
  $u_1$, foldcat dist. & 0.1  \\  
  $D$ & $0.01$ \\
  Implied $\alpha$ & 2.72  \\
  Fraction of folding & \\
  chains that are foldcats & 0.047 \\
 \hline
\end{tabular}
\caption{\textbf{Parameters used for Figs~\ref{fig:guseva-buckets} and \ref{fig:random-fold-foldcat}.} Units were normalized so that $u_1 = 1$ for the Flory distribution (determines concentration units) and $d(2) = 1$ for the Flory distribution (determines time units). Folding was chosen to peak around $n=5$, giving the bumps in the red and blue curves of Fig~\ref{fig:random-fold-foldcat}.} \label{tab:analytical-params}
\end{table}
\end{center}

Adding foldcats to our model of folding, Equation~\eqref{eq:random-poly-plus-folding}, is simple. Assume that a fraction $q_{fc}(n)$ of the folded polymers $f_n$ are foldcats. Then, the total number of foldcats is $C = \sum q_{fc}(n) f_n$. If foldcats catalyze the elongation of unfolded chains with a rate $k_c C u_1 u_n$, then we can replace the rate constant $k$ in our equations with the quantity $R = k + k_c C$. Our full model is then

\begin{align}
    \frac{du_1}{dt} &= \alpha - D u_1 - 2 R u_1^2 - \sum\limits_{n=2}^\infty R u_n u_1 \; , \nonumber \\
    \frac{du_n}{dt} &= R u_1 (u_{n-1} - u_n) - d(n) u_n + k_u(n) f_n - k_f(n) u_n \; , \label{eq:foldcat-full} \\
    \frac{df_n}{dt} &= k_f(n) u_n - k_u(n) f_n - d_f(n) f_n \; . \nonumber
\end{align}

We should once again look for the steady-state distribution. We cannot immediately proceed as before, because $R$ now depends on the entire distribution $u_n$. We no longer get a recursive solution. As before, however, we can recognize that the constants $q_{fc}(n)$ are not what we care about. We can instead swap them out for knowing the level of foldcats $C$. We have a valid solution for some $q_{fc}(n)$ as long as $C \leq \sum f_n$. Just as we traded the supply rate $\alpha$ for the more important parameter $u_1$, we now trade $q_{fc}(n)$ for the more important C. In doing so, we get the folding solution of Equation~\eqref{eq:pn-general-case-folding}, except with $k$ replaced by $R$. In principle, we could also solve this system of equations self-consistently; we do not do this analysis here. \\

To compare the foldcat model to the generalized Flory distribution with and without folding, we must simply make sure we have the same $\alpha$ for all three. This is done in Figure~\ref{fig:random-fold-foldcat}, which shows the fraction of chains at each length for random Founding Rock polymerization only (Flory distribution, black), folding (blue) and foldcats (red). The presence of foldcats allows for many orders of magnitude more long chains in the population. The total number of chains increases in the folding and foldcat cases as well. The parameters used to create this figure, which were just examples chosen to illustrate the foldcat effect, are shown in Table~\ref{tab:analytical-params}. These same parameters were used to create Fig~\ref{fig:guseva-buckets} in the main text.

\section{A binned version of the chain-length-only foldcat model is a two-step process driven by chain persistence.}\label{app:binned-model}

\renewcommand{\theequation}{B\arabic{equation}}
\setcounter{equation}{0}

Instead of using the full foldcat model of Eq~\eqref{eq:foldcat-full} to find the time-dependent behavior of the Foldcat Mechanism, we used a binned model. We tracked ten species: monomers $u_1$, then chains that do not fold at all $u$, chains that can fold $f$, and foldcats $c$, the three of which could each have lengths short $s$, medium $m$, and long $l$. Our categories were $u_1$, $u_s$, $u_m$, $u_l$, $f_s$, $f_m$, $f_l$, $c_s$, $c_m$, and $c_l$. To compensate for losing specific length information, the elongation reactions were now assumed to be probabilistic. For example, when a $u_s$ is elongated, there is some probability $p_{usus}$ that is stays a $u_s$, some probability $p_{usfs}$ that it becomes an $f_s$, some probability $p_{uscm}$ that it becomes a $c_m$, and so on. Elongation was assumed to be able to change the category to any other either at the same level (e.g. $s$ goes to $s$) or at the next higher level (e.g. $s$ goes to $m$ and $m$ goes to $l$). Of course, $\sum_j p_{kj} = 1$ is the normalization condition. \\

The full model, then, included nonequilibrium supply of monomer, decay of every species back into monomers using an average $n_k$ monomers recovered for decays of chains of length k, joining of two $u_1$ into a $u_s$, and the probabilistic elongation of each species that we described above. The ODEs for this binned model are

\begin{widetext}
\begin{align}
    \frac{d u_1}{dt} &= \alpha - D u_1 + n_s (d_{us}u_s + d_{fs} f_s + d_{cs} c_s) + n_m (d_{um}u_m + d_{fm} f_m + d_{cm} c_m) + n_l (d_{ul}u_l + d_{fl} f_l + d_{cl} c_l) - \nonumber  \\ &-(k_u + k_{uc}c_T)(u_s + u_m + u_l - 2 u_1) u_1 - (k_f + k_{fc}c_T)(f_s + f_m + f_l) u_1 - (k_c + k_{cc}c_T)(c_s + c_m + c_l) u_1 \; , \nonumber \\
    \frac{d u_s}{dt} &= (p_{usus} - 1) (k_u + k_{uc} c_T) u_s u_1 + p_{fsus}(k_f + k_{fc} c_T) f_s u_1 + p_{csus} (k_c + k_{cc} c_T) c_s u_1 - d_{us} u_s + (k_u + k_{uc} c_T) u_1^2 \; , \nonumber \\
    \frac{d u_m}{dt} &= p_{usum}(k_u + k_{uc} c_T) u_s u_1 + p_{fsum} (k_f + k_{fc} c_T) f_s u_1 + p_{csum} (k_c + k_{cc} c_T) c_s u_1 + \nonumber \\ 
    &+ (p_{umum} - 1)(k_u + k_{uc} c_T) u_m u_1 + p_{fmum} (k_f + k_{fc} c_T) f_m u_1 + p_{cmum}(k_c + k_{cc} c_T) c_m u_1 - d_{um} u_m \; , \nonumber \\
    \frac{d u_l}{dt} &= p_{umul} (k_u + k_{uc} c_T) u_m u_1 + p_{fmul} (k_f + k_{fc} c_T) f_m u_1 + p_{cmul} (k_c + k_{cc} c_T) c_m u_1 + \nonumber \\
    &+ (p_{ulul} - 1) (k_u + k_{uc} c_T) u_l u_1 + p_{flul} (k_f + k_{fc} c_T) f_l u_1 + p_{clul} (k_c + k_{cc} c_T) c_l u_1 - d_{ul} u_l \; , \nonumber \\
    \frac{d f_s}{dt} &= p_{usfs} (k_u + k_{uc} c_T) u_s u_1 + (p_{fsfs} - 1) (k_f + k_{fc} c_T) f_s u_1 + p_{csfs} (k_c + k_{cc} c_T) c_s u_1 - d_{fs} f_s \; , \label{eq:binned-model-full-ODEs} \\
    \frac{d f_m}{dt} &= p_{usfm} (k_u + k_{uc} c_T) u_s u_1 + p_{fsfm} (k_f + k_{fc} c_T) f_s u_1 + p_{csfm} (k_c + k_{cc} c_T) c_s u_1 + \nonumber \\ 
    &+ p_{umfm} (k_u + k_{uc} c_T) u_m u_1 + (p_{fmfm} - 1) (k_f + k_{fc} c_T) f_m u_1 + p_{cmfm} (k_c + k_{cc} c_T) c_m u_1 - d_{fm} f_m \; , \nonumber \\
    \frac{d f_l}{dt} &= p_{umfl} (k_u + k_{uc} c_T) u_m u_1 + p_{fmfl} (k_f + k_{fc} c_T) f_m u_1 + p_{cmfl} (k_c + k_{cc} c_T) c_m u_1 + \nonumber \\ 
    &+ p_{ulfl} (k_u + k_{uc} c_T) u_l u_1 + (p_{flfl} - 1) (k_f + k_{fc} c_T) f_l u_1 + p_{clfl} (k_c + k_{cc} c_T) c_l u_1 - d_{fl} f_l \; , \nonumber \\
    \frac{d c_s}{dt} &= p_{uscs} (k_u + k_{uc} c_T) u_s u_1 + p_{fscs} (k_f + k_{fc} c_T) f_s u_1 + (p_{cscs} - 1) (k_c + k_{cc} c_T) c_s u_1 - d_{cs} c_s \; , \nonumber \\
    \frac{d c_m}{dt} &= p_{uscm} (k_u + k_{uc} c_T) u_s u_1 + p_{fscm} (k_f + k_{fc} c_T) f_s u_1 + p_{cscm} (k_c + k_{cc} c_T) c_s u_1 + \nonumber \\ 
    & + p_{umcm} (k_u + k_{uc} c_T) u_m u_1 + p_{fmcm} (k_f + k_{fc} c_T) f_m u_1 + (p_{cmcm} - 1) (k_c + k_{cc} c_T) c_m u_1 - d_{cm} c_m \; , \nonumber \\
    \frac{d c_l}{dt} &= p_{umcl} (k_u + k_{uc} c_T) u_m u_1 + p_{fmcl} (k_f + k_{fc} c_T) f_m u_1 + p_{cmcl} (k_c + k_{cc} c_T) c_m u_1 + \nonumber \\
    & + p_{ulcl} (k_u + k_{uc} c_T) u_l u_1 + p_{flcl} (k_f + k_{fc} c_T) f_l u_1 + (p_{clcl} - 1) (k_c + k_{cc} c_T) c_l  u_1 - d_{cl} c_l \; , \nonumber 
\end{align}

\begin{center}
\begin{table}
\begin{tabular}{ ||c | c||c | c||c | c||c | c||c | c ||} 
 \hline
  $\alpha$ & 1 & $k_f$ & 0.008 & $p_{usus}$ & 0.864 & $p_{fsus}$ & 0.01 & $p_{csus}$ & 0.0055  \\  
  $n_s$ & 15 & $k_{fc}$ & 0.25 & $p_{usfs}$ & 0.10 & $p_{fsfs}$ & 0.718 & $p_{csfs}$ & 0.01 \\ 
  $n_m$ & 50 & $k_c$ & 0.008 & $p_{uscs}$ & 0.03 & $p_{fscs}$ & 0.27 & $p_{cscs}$ & 0.983 \\
  $n_l$ & 100 & $k_{cc}$ & 0.25 & $p_{usum}$ & 0.005 & $p_{fsum}$ & 0.0001 & $p_{csum}$ & 0.0002 \\  
  $k_u$ & 0.02 & $D$ & $5\times 10^{-5}$ & $p_{usfm}$ & 0.0007 & $p_{fsfm}$ & 0.0012 & $p_{csfm}$  &  0.0003 \\  
  $k_{uc}$ & 1 & Time units & $u_1(0)/\alpha$ & $p_{uscm}$ & 0.0003 & $p_{fscm}$ & 0.0007 & $p_{cscm}$ & 0.001 \\ 
  $p_{umum}$ & 0.718 & $p_{fmum}$ & 0.001 & $p_{cmum}$ & 0.0055 & $p_{ulul}$ & 0.5 & $p_{clul}$ & 0.0001 \\
  $p_{umfm}$ & 0.22 & $p_{fmfm}$ & 0.679 & $p_{cmfm}$ & 0.01 & $p_{ulfl}$ & 0.43 & $p_{clfl}$ & 0.07 \\  
  $p_{umcm}$ & 0.06 & $p_{fmcm}$ & 0.318 & $p_{cmcm}$ & 0.983 & $p_{ulcl}$ & 0.07 & $p_{clcl}$ & 0.9299 \\ 
  $p_{umul}$ & 0.0007 & $p_{fmul}$ & 0.0001 & $p_{cmul}$ & 0.0002 & $p_{flul}$ & 0.001 & Concentration units & $u_1(0)$ \\
  $p_{umfl}$ & 0.001 & $p_{fmfl}$ & 0.001 & $p_{cmfl}$ & 0.0003 & $p_{flfl}$ & 0.929 & &  \\
  $p_{umcl}$ & 0.0003 & $p_{fmcl}$ & 0.0009 & $p_{cmcl}$ & 0.001 & $p_{flcl}$ & 0.07 & &  \\ 
 \hline
\end{tabular}
\caption{\textbf{Parameters used for the binned foldcat model.} Values were determined using the heuristics described in the text. Units were normalized so that $u_1(0) = 1$ (concentration units) and $u_1(0)/\alpha = 1$ (time units).} \label{tab:params}
\end{table}
\end{center}
\end{widetext}

\noindent where $c_T = c_s + c_m + c_l$ is the total number of foldcats, $k_u$ is the rate of elongation of a $u$ (similarly for $k_f$ and $k_c$), and $k_{uc}$ is the rate of catalyzed elongation of a $u$ (similarly for $k_{fc}$ and $k_{cc}$). \\

There are some heuristics that we used to guide our choices of the values of the (very many) undetermined parameters of this model. Foldcats should be like folders in decay and elongation. It should be easier to elongate an unfoldable chain than a folder or foldcat - $k_u$ or $k_{uc}$ are bigger than their counterparts. On the flip side, unfoldable chains should decay faster. The probability of staying in the same category upon elongation should be highest, while the probability of going to the unfoldable category from either $f$ or $c$ should be small and decrease with chain length. The number of monomers released per chain at each size should be on the order of the previous category's average length plus the inverse of the probability of jumping up a category upon elongation. Our full list of constants used to make Fig~\ref{fig:guseva-persistence-ratchet}, which demonstrated the persistence-driven two-step nature of the random Founding Rock search followed by the directed foldcat search, is shown in Table~\ref{tab:params}. 

\section{Analyzing the limits of the invasion analysis approximation and the eventual resource limitation in various models.}\label{app:invasion-analysis}

\renewcommand{\theequation}{C\arabic{equation}}
\setcounter{equation}{0}

Our first objective is to see how the invasion analysis that produces Fig~\ref{fig:three-options} eventually reaches a stable steady-state population of autocatalysts (autocats) because of resource limitations. The starting point will be the resource competition equations 

\begin{align}
    \frac{dr}{dt} &= \alpha - d_r r - k(A) A r \; , \nonumber \\
    \frac{dA}{dt} &= k(A) A r - D A \; , \label{eq:DEM}
\end{align}

\noindent for one autocat $A$ on one resource $r$, as developed in, e.g., \cite{tilmanResourceCompetitionCommunity1982,kocherDarwinianEvolutionDynamical2023,kocherOriginsLifeFirst2023,macarthurSpeciesPackingCompetitive1970}. A full resource-dependent population model like the one in Eq~\eqref{eq:DEM} is the most basic starting point for any ``first invasion'' analysis. The resource $r$ is supplied at a rate $\alpha$, decays with rate constant $d_r$, and is eaten by the autocats with rate constant $k(A)$. The autocats decay with rate $D$. Setting the resource to steady-state, $dr/dt = 0$, and plugging back into the differential equation for $A$ gives

\begin{align}
    \frac{dA}{dt} &= k(A) A \frac{\alpha}{d_r + k(A) A} - D A \nonumber \\
    &= \frac{(\alpha k(A)/d_r) A}{1 + (k(A)/d_r) A} - D A \; . \label{eq:DEM-no-r}
\end{align}

\noindent Note that setting the resource to steady-state does not contradict the separation of timescales between the system's dynamics and environmental changes that we referred to in the text: we only assume that $\alpha$ and $d_r$ are slowly varying functions of time, not necessarily that they are slow compared to the autocat's dynamics (that is, our assumption was on their derivative, not their value). It is generally expected that, to get the single ODE for $A(t)$ needed for the invasion analysis, resources will be put to their instantaneous steady-state values. We now define $k_1 + k_2 A = (k(A) \alpha/d_r)$ by Taylor expanding $k(A)$. While we expanded $k(A)$ to second order here, there is no reason why $k_2$ should be non-zero. In each individual model, it must first be argued that the cooperativity exists, and a mechanism for it must be given, before taking $k_2 \neq 0$. This discussion must take place before applying the invasion analysis. Our ODE for the autocat population in this case is

\begin{equation}
    \frac{dA}{dt} = \frac{k_1 A + k_2 A^2}{1 + (k_1 A + k_2 A^2)/ \alpha} - D A \; . \label{eq:DEM-no-r-units}
\end{equation}

\noindent We can see that the parameter $\alpha$, which characterizes the nonequilibrium driving of the system by resource supply, determines when the invasion analysis is valid. When $\alpha \gg k_1 A + k_2 A^2$, we can Taylor expand Eq~\eqref{eq:DEM-no-r-units} to find exactly Eq~\eqref{eq:templated-polymerization-type} with $g_1 = k_1$ and $g_2 = k_2 - k_1^2 / \alpha$. This is only one example; the actual formula for the parameter that determines when the invasion analysis is valid, or for $g_1$ and $g_2$, will depend on the specific mechanism involved (the interplay between the autocatalyst and the resource, as well as the nonequilibrium driving). The Foldcat Mechanism model of Eq~\eqref{eq:foldcat-type} from the main text, for example, has the condition that $A \ll A_s$ and $k_1 A \ll 1$, and we will give further examples shortly. To move to unitless variables, we take time to be measured in units of $1/D$ and concentration to be measured in units of $\alpha/D$. Our unitless parameters would be $\Tilde{A} = D A/\alpha$, $\tilde{k}_1 = k_1 / D$, and $\Tilde{k}_2 = k_2 \alpha / D^2$. Dropping the tildes, we find:

\begin{equation}
    \frac{dA}{dt} = \frac{k_1 A + k_2 A^2}{1 + k_1 A + k_2 A^2} - A \; . \label{eq:DEM-no-r-unitless}
\end{equation}

The invasion analysis limit is now $(k_1 A + k_2 A^2) \ll 1$. The corresponding potential landscape is 

\begin{align}
    V(A)& = \frac{A^2}{2} - A - \nonumber \\ 
    -&\frac{2}{\sqrt{4 k_2 - k_1^2 }} \Biggl(\arctan\left[ \frac{k_1}{\sqrt{4 k_2 - k_1^2}} \right] -  \nonumber \\
     -&\arctan\left[ \frac{k_1 + 2 A k_2}{\sqrt{4 k_2 - k_1^2}} \right] \Biggr) \; . \label{eq:success-catastrophe-fitness}
\end{align}

\begin{figure*}[t]
    \centering
    \includegraphics[width = 0.6\linewidth]{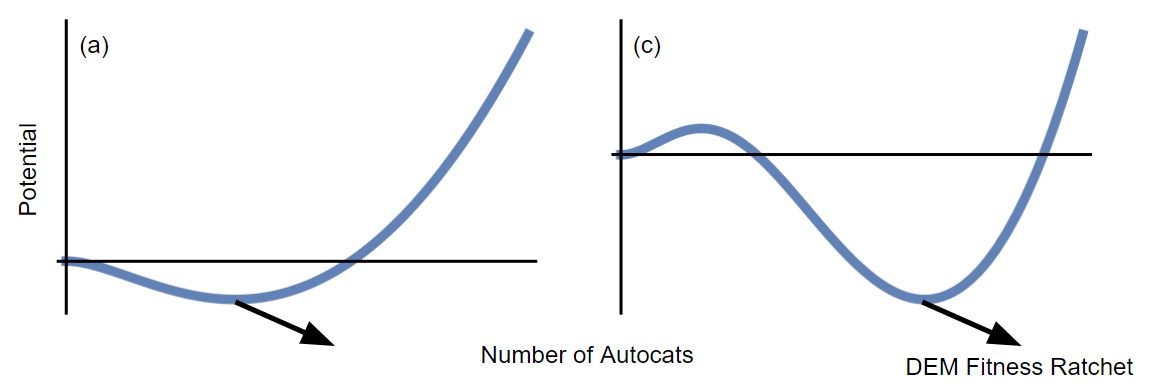}
    \caption{\textbf{Growth stops when resources are fully subscribed, but then evolution takes over.} Situations (a) and (c) as in Figure~\ref{fig:three-options}, but now with resource limitations included. The Fig~\ref{fig:three-options} is the small $A$ limit of this one. Each population saturates at a persistent, finite value, at which point evolution acts (through e.g. the Darwinian evolution machine, the DEM \cite{kocherDarwinianEvolutionDynamical2023}) to ratchet up the fitness (black arrow pointing down and to the right).}
    \label{fig:finite-populations}
\end{figure*}

\begin{figure}[t]
    \centering
    \includegraphics[width = \linewidth]{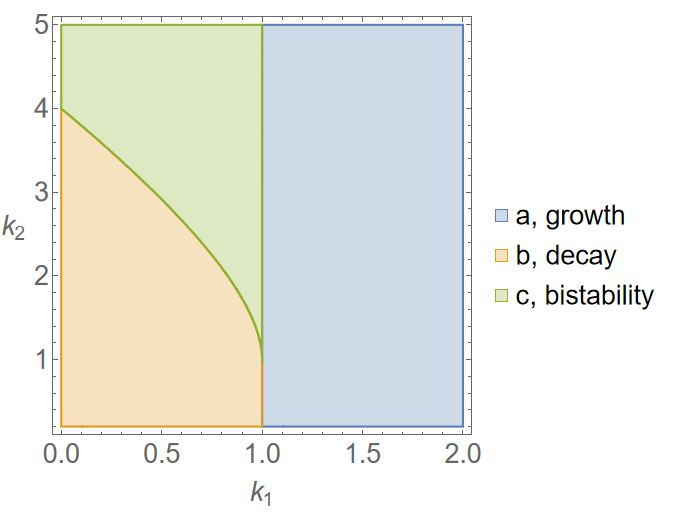}
    \caption{\textbf{Phase diagram of cases (a) growth, (b) decay, and (c) bistability} when varying the two dimensionless parameters $k_1$, which is the rate of non-cooperative reproduction, and $k_2$, which is the rate of cooperative reproduction, for the potential landscape Eq~\eqref{eq:success-catastrophe-fitness}.}
    \label{fig:success-catastrophe-phase-diagram}
\end{figure}

\noindent For cases (a) and (c) as defined in Fig~\ref{fig:three-options}, we get the full resource-limited dynamics in Fig~\ref{fig:finite-populations}. In each case, the population settles to a persistent level when the resource is fully utilized. When $A$ has a perisistent population, evolution acts to increase the fitness (move to higher population and lower potential). The invasion analysis figures are the zoomed-in portion near $A = 0$. Case (b) can be considered to be a part of the case (c) plot of Fig~\ref{fig:finite-populations} if the potential barrier does not occur within the invasion analysis regime (our next analysis). The phase diagram for this model is shown in Fig~\ref{fig:success-catastrophe-phase-diagram}. \\ 

Now that we know when the invasion analysis is valid, we turn to finding when the autocatalyst would be cooperative in the invasion analysis regime. To do this, we need to find the potential barrier maximum of Eq~\eqref{eq:templated-polymerization-type} and compare it to the parameter that determines when the invasion analysis is valid. In general, we will call this parameter $B$ (for the previous model, we found $B = \alpha$; for foldcats, $B = \max\{A_s,1/k_1\}$). If the minimum is greater than $B$, then the autocatalyst is not cooperative: the only way to fluctuate into the funnel of the persistent state is to leave the invasion analysis regime (presumably a very rare event because of the size of fluctuations needed). We will not plug in explicitly for $B$; instead, we will only use the parameters defined in the main text. Our starting point is setting Eq~\eqref{eq:templated-polymerization-type} equal to zero. The minimum location is 

\begin{equation}
    A_0 = \frac{D - g_1}{g_2} \; . \label{eq:A0-explicitly}
\end{equation}

\noindent For our model, then, a cooperative autocat is defined by $A_0 = \frac{D - g_1}{g_2} \lesssim B$, where $B$ is found using the specific mechanism as we illustrated above. \\

If we are looking at a cooperative autocatalytic process with two separate resources put to steady-state for the linear and quadratic ``birth'' terms, then instead of Equations~\eqref{eq:DEM-no-r-unitless} and \eqref{eq:success-catastrophe-fitness} we could use 

\begin{equation}
    \frac{dA}{dt} = \frac{k_1 A}{1 +  A / B} + \frac{k_2 A^2}{1 +  A^2 / C^2} - D A\; , \label{eq:full-saturation-TP}
\end{equation}

\noindent with corresponding potential landscape 

\begin{align}
    V(A) = \frac{D A^2}{2} - (k_1 B &+ k_2 C^2) A + k_2 C^3 \tan^{-1}\left(\frac{A}{C} \right) - \nonumber \\ 
    &- k_1  B^2 \ln\left( \frac{B}{B + A} \right) \; . \label{eq:success-catastrophe-fitness-TP}
\end{align}

\noindent Now, cooperativity in the invasion analysis regime would be characterized by $A_0 \lesssim B$ and $A_0 \lesssim C$. Both constraints would have to be satisfied for an autocatalyst to be considered case (c). \\

Finally, we will consider a templated polymerization model. We have templated polymerizers $A$, monomers $M$, and non-catalytic polymers $r$. The reactions are $A + r + M \rightarrow 2 A + r$ (incorrect copying of an inactive chain to make an active chain) and $2 A + M \rightarrow 3 A$ (correct copying of an active chain to make another active chain), $A + r + M \rightarrow A + 2 r$ (correct copying of a non-autocatalytic chain), and $2A + M \rightarrow 2 A + r$ (incorrect copying of an autocatalytic chain). Note that the second of the reactions, producing $3A$, is the cooperative one (e.g. consider its mass-action rate, which is proportional to $A^2$). The full ODE model of these reactions, including the nonequilibrium-driven supply terms\footnote{In general, $\alpha_r$ should be a complicated function of $M$. We will ignore that. It actually does not affect the invasion analysis that we do here, since $M$ is approximately constant anyway. We assume the use of monomers to make $r$ is rolled into the decay term $d_M$.} $\alpha_r$ and $\alpha_M$, would be 

\begin{align}
    \frac{dr}{dt} &= \alpha_r - d_r r + k_1 A r M + k_4 A^2 M \; , \nonumber \\
    \frac{dM}{dt} &= \alpha_M - d_M M - (k_1 + k_2) A r M - (k_3 + k_4) A^2 M \; , \label{eq:TP-ODE} \\
    \frac{dA}{dt} &= k_2 A r M + k_3 A^2 M - DA \; , \nonumber 
\end{align}

\noindent where $k_1$ is the mass-action rate of faithful copying of an $r$, $k_2$ is the rate of unfaithful copying of an $r$, and $k_3$ and $k_4$ are the rates of faithful and unfaithful copying of an $A$, respectively. By assuming $M$ and $r$ attain their steady-states with $A = 0$, we can find that the maximum of the potential barrier for the templated polymerizers would appear at 

\begin{equation}
    A_0^{TP} = \frac{d_r d_M D - k_2 \alpha_r \alpha_M}{k_3 \alpha_M d_r} \; . \label{eq:A0-templated-polymerizer}
\end{equation}

We could proceed as previously, putting the resources $r$ and $M$ to steady-state then comparing $A_0^{TP}$ to the concentration parameter that determines when we can Taylor expand the rate equation for $A$, but the expressions for $r_{SS}$ and $M_{SS}$ are unwieldy (although we should note that if $r$ is presumed to be constant, and the processes $k_1$ and $k_4$ are ignored, then Eq~\eqref{eq:TP-ODE} reduces to Eq~\eqref{eq:DEM-no-r-unitless}). Instead, we will approximate. We note that the invasion analysis approximation is when $r \approx \alpha_r/d_r$ and $M \approx \alpha_M / d_M$. That is, only the zeroth order term of the Taylor expansion of $r$ and $M$ in powers of $A$ contributes. Explicitly, this expansion is $r = \alpha_r / d_r + g_r A$ and $M = \alpha_M / d_M + g_M A$ for some constants $g_r$ and $g_M$. The invasion analysis approximation holds at the potential barrier maximum if $g_r A_0^{TP} d_r / \alpha_r \lesssim 1$ and $g_M A_0^{TP} d_M / \alpha_M \lesssim 1$. Setting the first two ODEs of Equation~\eqref{eq:TP-ODE} to steady-state using only terms up to linear order in $A$ gives 

\begin{align}
    g_r &= \frac{k_1 \alpha_r \alpha_M}{d_r^2 d_M} \; , \nonumber \\
    g_M &= \frac{(k_1 + k_2) \alpha_r \alpha_M}{d_M^2 d_r} \; , \label{eq:grgM}
\end{align}

\noindent so that the conditions for a cooperative autocat are

\begin{align}
    \frac{k_1 \alpha_M A_0^{TP}}{d_r d_M} &\lesssim 1 \; , \nonumber \\
    \frac{(k_1 + k_2) \alpha_r A_0^{TP}}{d_r d_M} &\lesssim 1 \; . \label{eq:TP-coop}
\end{align}

\section{The foldcat-like cooperativity model.}\label{app:foldcat-model}

\renewcommand{\theequation}{D\arabic{equation}}
\setcounter{equation}{0}

First, we should see why foldcat catalysis is actually cooperative in the first place. For templated polymerizers, it is easy to see that the reaction $2 A + M \rightarrow 3A$ has a mass-action rate proportional to $A^2$, giving the cooperativity. On the other hand, foldcats catalyze elongation reactions. One such reaction would be a foldcat $A$ taking a non-catalytic chain $r$ and adding a monomer $M$ to it to make a resultant chain that is another foldcat: $A + r + M \rightarrow 2A$. Alternatively, a foldcat can elongate a foldcat and keep it as a foldcat, $2A + M \rightarrow 2A$. In the latter reaction, there is no net change in the number of foldcats. In the former, one foldcat is added, but the mass-action rate of reaction is only proportional to $Ar$. At first glance, then, it seems like foldcats are not cooperative, since there is no factor of $A^2$. However, foldcats also catalyze the formation of $r$ from only $M$. Thus, the amount of precursor $r$ should be a function of $A$; Taylor expanding $r(A) = r_0 + r_1 A$ gives the extra factor of $A$ necessary for cooperativity in foldcat dynamics. \\

We will now write down a full toy model of the Foldcat Mechanism just as we did for templated polymerization in the previous Appendix section. We will then simplify this model to the form that we worked with in the main text. Our basic reactions are that monomer $M$ is nonequilibrium supplied at a rate $\alpha_M$ and decays at a rate $d_M M$, while non-catalytic chains $r$ are created at a rate $\alpha_r$ and decay at a rate $d_r r$. Then, we have the elongation reactions, which are catalyzed by the founding rock and by our foldcats $A$: (1) $r + M \rightarrow A$ (non-foldcat is elongated into a foldcat), (2) $r + M \rightarrow r$ (non-foldcat elongated and still is not a foldcat),  (3) $A + M \rightarrow A$ (foldcat is elongated and is still a foldcat), and (4) $A + M \rightarrow r$ (foldcat is elongated and is no longer a foldcat). Elongation reaction $i$ has mass-action rate constant $K_i(A)$. The full ODE model is

\begin{align}
    \frac{dr}{dt} &= \alpha_r - d_r r + K_4(A) A M - K_1(A) r M \; , \nonumber \\
    \frac{dM}{dt} &= \alpha_M - [d_M + r (K_1(A) +  K_2(A)) + \nonumber \\ 
    &+ A (K_3(A) +  K_4(A))] M \; , \label{eq:foldcat-type-full-system-ODE} \\
    \frac{dA}{dt} &= K_1(A) r M - K_4(A) A M - D A \; . \nonumber 
\end{align}

Foldcats also cause the creation of chains $r$ from just monomer, so $\alpha_r = \alpha_r(A,M)$. This observation\footnote{In a more detailed model that kept track of sequence as well, the important point is that foldcats create sequences that can be elongated into foldcats, not just that foldcats create dimers from monomer as here.} is key for the cooperativity of foldcats. If we now suppose that $M$ is a constant everywhere except in the function $\alpha_r(A,M)$, where we assume it takes on a saturating form as the resource in Eq~\eqref{eq:DEM-no-r} does, we have the reactions 

\begin{align}
    \frac{dr}{dt} &= \alpha + \frac{\alpha_2 A}{1 + A/A_s} - d_r r + K_4'(A) A - K_1'(A) r \; ,  \label{eq:foldcat-type-no-M} \\
    \frac{dA}{dt} &= K_1'(A) r - K_4'(A) A  - D A \; , \nonumber 
\end{align}

\noindent where we have absorbed the constant amount of monomer $M$ into the definitions of the functions $K_i'(A) = M K_i(A)$. The quantity $A_s$ determines when the foldcats' creation of $r$ can go no faster. We are most concerned with foldcat untethering from the Founding Rock, so we make some further assumptions. Founding Rock terms in $K_i'(A)$ would be constants, while foldcat terms are, in the first approximation, proportional to $A$. We drop the constant terms. Further, we assume the reaction (4) that elongates foldcats into non-foldcats is very rare, so we drop it altogether. Upon defining $\alpha_2 = \alpha k_2$, we find the ODEs

\begin{align}
    \frac{dr}{dt} &= \alpha + \frac{\alpha k_2 A}{1 + A/A_s} - d_r r - K_1' A r \; ,  \label{eq:foldcat-type-simplified} \\
    \frac{dA}{dt} &= K_1' A r  - D A \; . \nonumber 
\end{align}

\noindent Setting the non-catalytic chains $r$ to steady-state finally gives the dynamical equation for the foldcats:

\begin{equation}
    \frac{dA}{dt} = \left( \alpha + \frac{\alpha k_2 A}{1 + A/A_s} \right) \frac{K_1' A}{d_r + K_1' A} - D A \; . \label{eq:foldcat-type-no-r}
\end{equation}

\noindent Further simplification is now possible by redefining $k_1 = K_1' / d_r$, which eliminates the parameter $d_r$, and by choosing units in which $\alpha = 1$ and $D = 1$. This choice is equivalent to measuring time in units of $1/D$ and measuring concentration in units of $\alpha / D$. Finally, we are left with the rate equation considered in the main text in terms of only dimensionless variables:

\begin{equation}
    \frac{dA}{dt} = \frac{k_1 A}{1 + k_1 A} + \frac{k_1 k_2 A^2}{(1 + k_1 A)(1 + A/A_s)} - A \; . \label{eq:foldcat-type-appendix-version}
\end{equation}

\noindent The parameters of our model are $k_1$, which is related to the rate at which foldcats elongate non-catalytic chains into foldcats; $k_2$, which is related to the rate at which foldcats create their direct precursors (a foldcat minus one monomer) from monomers or shorter non-catalytic chains; and $A_s$, which is the Michaelis constant of the creation of direct precursors from monomers or shorter non-catalytic chains. \\

To include the cooperative effects of increased chain stability, we can change the decay term as indicated in the main text, slowing the decay by a new term that has saturation constant $B_s$. The full ODE is 

\begin{equation}
    \frac{dA}{dt} = \frac{k_1 A}{1 + k_1 A} + \frac{k_1 k_2 A^2}{(1 + k_1 A)(1 + A/A_s)} - \left(A - \frac{k_3 A^2}{1 + A^2/B_s^2}\right) \; , \label{eq:foldcat-type-with-decay-cooperativity}
\end{equation}

\noindent where the term in parentheses is constrained to be positive (this requirement puts a constraint on $k_3$ and $B_s$, namely that $B_s^2 k_3^2 < 4$). The full potential function for foldcats with both types of cooperativity is  

\begin{align}
    V(A) = &\frac{A^2}{2} - A - A A_s k_2 - \frac{A_s^3 k_1 k_2 \ln(1 + A/A_s)}{1 - A_s k_1} - \nonumber \\
    & - \frac{(1 - A_s k_1 + A_s k_2) \ln(1 + k_1 A)}{k_1(A_s k_1 - 1)} -  \nonumber \\
    & - A B_s^2 k_3 + B_s^3 k_3 \arctan\left(\frac{A}{B_s}\right)\; .  \label{eq:foldcat-type-fitness-both-cooperativities}
\end{align}







\bibliography{refs}

\end{document}